# Magnetic shielding in MgB$_2$/Fe superconducting wires

J. Horvat, S. Soltanian, X. L. Wang, and S. X. Dou

*Abstract*—Transport critical current (I$_c$) was measured for MgB$_2$/Fe round wires, with magnetic field oriented perpendicular to the wire and parallel to it. Measurements were made on a wire with pure MgB$_2$ core and another wire where MgB$_2$ core was doped with nano-size SiC. This doping strongly improved the vortex pinning in MgB$_2$. The field dependence of I$_c$ was strongly improved due to the presence of the iron sheath. At 30K, I$_c$ did not depend on the field for fields between 0.09 and 0.7T. At lower temperatures, I$_c$ increased with the field, after initial decrease, resembling a "peak effect". This effect was extended to higher fields as the temperature went down. This improvement was not due to mere magnetic shielding by iron, but more likely to an interaction between the iron sheath and superconductor. Improvement of vortex pinning did not affect the range of fields within which this effect was observed.

## I. INTRODUCTION

MgB$_2$ superconductor is a new candidate for making superconducting wires that can be used without the need for liquid helium. Its relatively high critical temperature of 39K makes it possible to use it at 20K, a temperature readily available with the modern cryo-coolers. Quite early in the development of MgB$_2$ wires, it became clear that iron is one of the best materials for the use as a sheath for the wires [1]-[4]. Transport critical current density (J$_c$) of $1.3 \times 10^5$ A/cm$^2$ was obtained at 20K and 1.7T for the samples measured here. In addition to providing a medium for obtaining MgB$_2$ in chemical reaction at high temperature and ensuring the mechanical strength of the wires, iron is a ferromagnetic material which can be utilized to magnetically shield the superconductor from the external field. This could be very effectively employed for decoupling the superconducting filaments in a multifilamentary MgB$_2$/Fe wires, substantially lowering the AC loss in the wires.

Our first results on the influence of iron sheath on the field dependence of transport J$_c$ revealed a much better improvement of J$_c$ than expected from a mere magnetic shielding [5]. There was a range of the fields where J$_c$ did not change with the field. This was attributed to the interaction between the superconductor and iron sheath, because it was measured at high fields, for which the magnetic shielding was no longer effective. This range of fields was found to widen with lowering the temperature, indicating that it may be possible to obtain a very weak field dependence of J$_c$ at 20K, at high fields. However, the lowest temperature at which the measurements could be performed was 32K, which was limited by the maximum current available from the pulse current source. Here, we were using an improved current source, enabling measurements at temperatures less than 20K. Improving the vortex pinning by newly discovered nano-SiC doping [6], we also test how the vortex pinning affects this effect.

## II. EXPERIMENTAL PROCEDURE

Superconducting wires were prepared by filling the iron tube with a mixture of 99% pure powders of magnesium and boron. For one of the wires, 10 wt.% of amorphous nano-size SiC was added. The wires were drawn to a diameter 1.4mm, with the diameter of the inner core of 0.8mm. Heating in flowing argon to 800°C for 15 minutes, MgB$_2$ was formed in a chemical reaction. As shown recently, addition of SiC strongly improved the vortex pinning of MgB$_2$, resulting in much improved field dependence of critical current density (J$_c$) [6]. The details of the wire preparation can be found elsewhere [7]. The wires measured had the same outer diameter and diameter of superconducting core. The sheath was made from the same iron tube for all the wires, ensuring the magnetic properties of the sheath were the same for all the samples measured.

Because critical current for these wires was hundreds of amperes, the transport measurements had to be performed by a pulse-method, to avoid heating. A pulse of current was obtained by discharging a capacitor through the sample, coil of thick copper wire and non-inductive resistor connected in series. The current was measured via the voltage drop on the non-inductive resistor of 0.01 Ohm. With a proper choice of the coil, the current reached its maximum value (700A) within 1ms. The voltage developed on the sample was measured simultaneously with the current, using a 2-channel digital oscilloscope. Because both channels of the oscilloscope

Manuscript received August 8, 2002. This work was supported by Australian Research Council and Hyper Tec Research Inc., OH, USA.

Authors are with Institute for Superconducting and Electronic Materials, University of Wollongong, NSW 2522, Australia (phone: -61-2-4221 5722; fax: -61-2-4221 5731; e-mail: jhorvat@uow.edu.au).

had the same ground, the signal from the voltage taps was first fed to a transformer preamplifier (SR554). This decoupled the voltage taps from the resistor used for measuring the current, thereby avoiding creation of the ground loops and parasitic voltages in the system, as well as of an additional current path in parallel to the sample. The transformer amplified the voltage 100 times, improving the sensitivity of the experiment.

A typical V-I characteristic is shown in Fig.1. Self-field of the pulse of the current induced a voltage in the voltage taps, which gave a background voltage. This voltage increased gradually at the beginning of the pulse, but was almost constant for the most of the pulse duration (Fig.1). It was easy to distinguish the voltage created by the superconductor on this background, because the voltage developed very abruptly when the current reached the value of $I_c$.

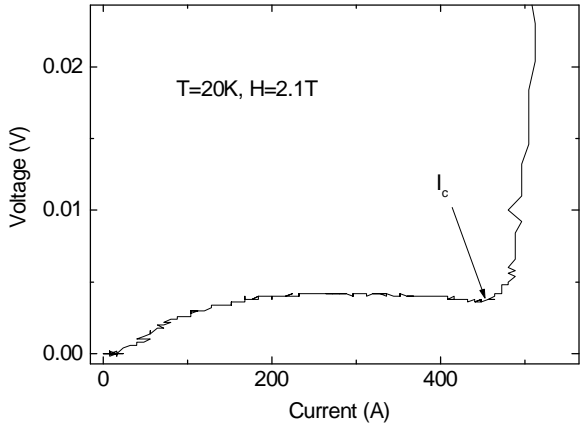

Fig. 1: A typical voltage-current characteristic for MgB$_2$/Fe wire obtained by the pulse-method.

Magnetic field was produced by a 12T superconducting magnet. Sample mounting allowed for orienting the field either perpendicular to the wire, or parallel to it. In the later case, the field was also parallel to the current passing through the sample. The sample was placed into a continuous flow helium cryostat, allowing the control of temperature better than 0.1K.

### III. EXPERIMENTAL RESULTS

Figure 2 shows the field dependence of $J_c$ for a non-doped wire at T=32, 30, 27, 24 and 22K. For all the temperatures, there is an initial decrease of $J_c$ with the field, up to a field $H_0$. After the decrease, $J_c$ is almost independent of the field at T=30 and 32K, up to a value of field $H_p$. For $H>H_p$, $J_c$ decreases with H exponentially. For lower temperatures, $J_c$ starts increasing with the field for $H>H_0$, giving a field dependence of $J_c$ resembling a "peak effect" (Fig.2). $J_c$ peaks at $H=H_p$ and it decreases exponentially with field for $H>H_p$ (Fig.2). The occurrence of the plateau and peak in $J_c$ vs. H improves the field dependence of $J_c$, as shown previously for T>30K [5]. This was shown to occur because of the interaction between the Fe sheath and superconductor. The field range with improved $J_c$ widens with decreasing temperature. For lower temperatures, the critical current at low fields was higher than the maximum current available with our experimental set-up and the peak could not be measured. Qualitatively the same results were also obtained for the SiC doped wire, which had improved field dependence of $J_c$ for $H>H_p$.

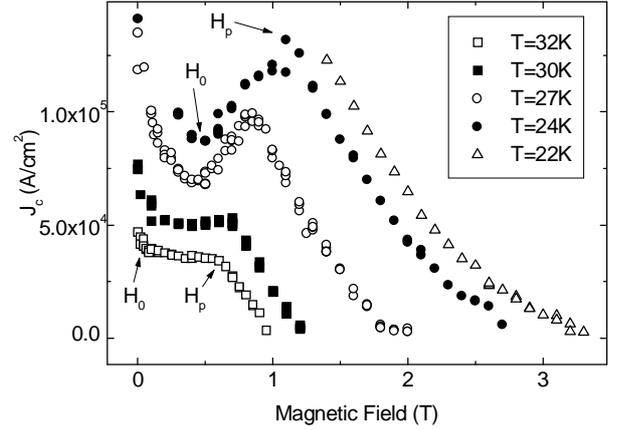

Figure 2: Field dependence of $J_c$ for undoped MgB$_2$ wire, with field perpendicular to the wire.

To study the temperature dependence of the field range with improved $J_c$, $H_p$ is plotted as a function of temperature in Fig.3. It should be noted that $J_c$ is improved by Fe sheath even for fields higher that $H_p$ for T<30K (Fig. 2), however there is no clear feature in $J_c$ vs. H for $H>H_p$ to enable a clear distinction of the field range with improved $J_c$. As shown in Fig. 3, $H_p$ decreases with temperature exponentially:

$$H_p = 7.61 Tesla \exp\left(-\frac{T}{12.36 Kelvin}\right) \quad . \quad (1)$$

The temperature dependence of $H_p$ is shown in Fig. 3 by solid and open symbols for pure and SiC doped wires, respectively. Apparently, improvement of vortex pinning by doping does not affect the values of $H_p$.

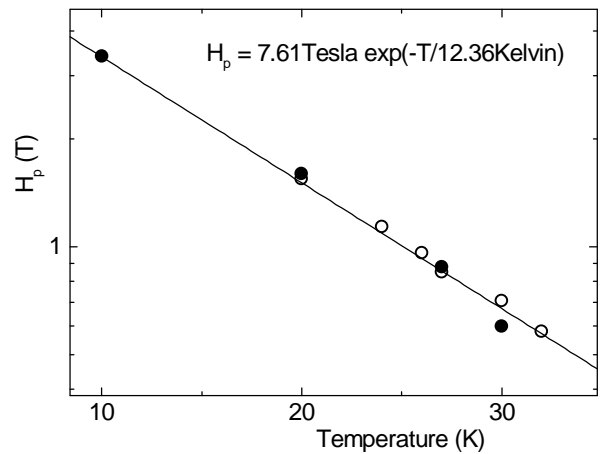

Fig. 3: Temperature dependence of $H_p$ for undoped (open symbols) and SiC doped (solid symbols) MgB$_2$/Fe wires. Solid line is the fit with Eq. (1).

Figure 4 shows the field dependence of $J_c$ for a pure $MgB_2$/Fe wire at 30K, with field parallel to the wire (open symbols), as compared to the field perpendicular to the wire (solid symbols). As opposed to the perpendicular field, $J_c$ for the parallel field initially increases with the field, up to a field $H_1$. This is followed by an exponential decrease of $J_c$ with field for $H>H_1$. At lower temperatures (below 27K), $J_c$ is almost constant for $H<H_1$. The value of $H_1$ decreases with temperature as: $H_1 = 3303$ Tesla exp(-T/ 3 Kelvin ) (Fig.5), much faster than $H_p$.

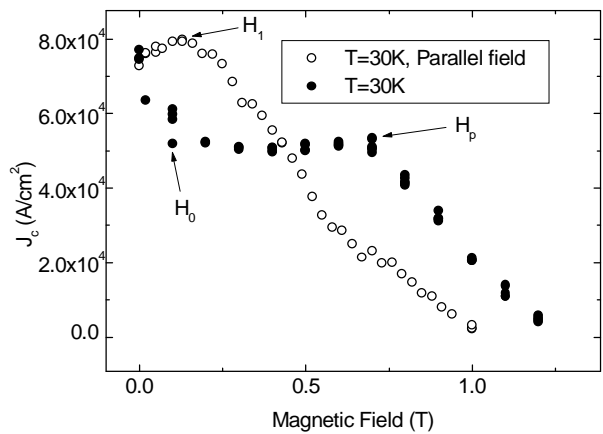

Fig.4: Field dependence of $J_c$ for an undoped $MgB_2$/Fe wire at 30K, with field perpendicular (solid symbols) and parallel (open symbols) to the wire.

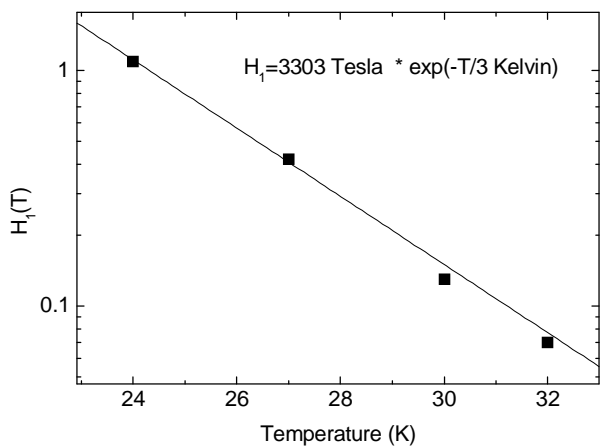

Fig.5: Temperature dependence of $H_1$ for an undoped $MgB_2$/Fe wire, with field parallel to the wire. Solid line is the fit with the exponential function.

IV. DISCUSSION

The plateau in the field dependence of $J_c$ has been reported earlier for $MgB_2$/Fe wires, for T>30K [5]. The plateau was observed to widen with decreasing temperature. It was suggested that the plateau occurred as a consequence of the interaction between the superconductor and iron sheath. In this work, we investigated the temperature dependence of the plateau for T<30K, thanks to an improved pulse current source enabling measurements of higher values of $I_c$.

The plateau was observed for T>30K in this work, as well. However, instead of the plateau, an increase of $J_c$ with H was observed for T<30K, resulting in a peak in the field dependence of $J_c$ (Fig.2). The peak apparently develops from the plateau as the temperature decreases, because the temperature dependence of $H_p$ follows the same exponential law for T>30K and T<30K. Measurements of magnetic shielding of hollow Fe cylinder showed that the shielding is effective up to about 0.2T [5]. The effectiveness of the shielding should not vary strongly at temperatures between 20 and 32K, because the Curie temperature of iron is much higher than this. Apparently, it is the change of superconducting properties of $MgB_2$ core that is responsible for the exponential decrease of $H_p$ with T. Because of this, strong improvement of vortex pinning with SiC doping was expected to affect the temperature dependence of $H_p$. To our surprise, this did not occur (Fig. 3).

The field dependence of $J_c$ with the field parallel to the wire is also influenced by the iron sheath. It was shown earlier that there is a plateau in $J_c(H)$ for H<0.02T and T=32K [5]. This was interpreted by simple magnetic shielding of the iron sheath. Namely, measurements of the field inside the shield for the parallel field showed that the internal field was almost zero and constant for H<0.02T. All of the field in excess to 0.02T penetrated the iron sheath for this field configuration. Here, we show that for T<32K, $J_c$ slightly increases with the field (Fig. 4) and the field range in which this occurs (H<$H_1$) increases strongly with lowering the temperature (Fig.5). The field dependence of $H_1$ is actually much stronger than of $H_p$. In addition, it was shown earlier that the magnetic shielding for the parallel field is very weak [5]. Again, such strong temperature dependence of $H_1$ cannot be ascribed to a simple magnetic shielding. Apparently, the interaction between the superconducting core and iron sheath is responsible for the anomalous improvement of the field dependence of $J_c$ for both, parallel and perpendicular field.

Improvement of critical current of a superconductor via interaction with its magnetic surrounding was studied theoretically by Genenko et al. [8]. In their model, critical currents were calculated for type II superconducting thin strips in partly filled vortex state, in magnetic surrounding. Only self-field of the transport current was considered and the magnetic surrounding was supposed to be reversible, linear and homogeneous. They considered the influence of magnetic surrounding on the current distribution in the central vortex-free part of the strip. The current distribution was found to be very sensitive to the shape of the magnetic surrounding and its distance to the strip. They predicted an increase of the maximum loss-free current by a factor of 100 when a superconducting strip is placed in an open magnetic cavity. However, for the smallest practically achievable distance between the superconductor and magnetic surrounding of 0.1mm, the maximum loss-free current is predicted to be 7 times larger than the critical current without the magnetic surrounding. This increase of the apparent critical current occurs because of distribution of the current into the vortex-free region of the superconducting strip, whilst the vortices are confined to the edges of the strip. So far, no experiments were reported to verify this model.

In the model of Genenko et al. [8], the vortices were confined to the edges of thin superconducting strip by edge barriers. The model also considered the case of partly flux filled thick superconducting strip, with weak edge barriers and strong vortex pinning. Vortex pinning defined the critical current, according to the critical state model [9]. However, due to the pinning, the transport current is confined to the edges of the superconductor, where there is a constant gradient in the density of magnetic vortices defined by the pinning, and $J_c$ is constant. The model shows that the magnetic surrounding causes a distribution of the transport current into the flux free region of the superconductor, resulting in a loss-free current density exceeding the critical current defined by the critical state model [8]. Because of this, we expected to observe a change in the value of $H_p$ with improvement of the pinning by SiC doping, because the flux profile in the superconductor changes with the change of the vortex pinning. However, no change in $H_p$ was observed (Fig. 3).

The reason we did not observe the change in $H_p$ with SiC doping may be that the model was devised only for the self-field, whereas in our experiment we used additional external field. The geometry of our sample (round wire) was also not the same as the one in the model (strip). Because the model did not consider the case of external field, it is not clear if the change of the pinning should result in a change of $H_p$ in the model (even though intuitively, one would expect this to occur).

## V. Conclusion

It was found that, instead of the expected plateau in the field dependence of $J_c$, the value of $J_c$ increases with field at low temperatures, resembling a peak effect. The field of the peak in $J_c(H)$ increases exponentially with lowering temperature. Thanks to the peak, the value of $J_c$ at 20K and 1.7 T is almost the same as the zero-field $J_c$ at that temperature, of the order of $10^5$ A/cm$^2$. At still lower temperatures, the peak extends to higher fields, reaching 3.5T at 10K. Even though its exact origin is still not clear, this effect can be employed for improvement of the field dependence of $J_c$ of MgB$_2$ wires.